
\input harvmac.tex
\overfullrule=0mm

%
\def\frac#1#2{\scriptstyle{#1 \over #2}}

\def\pd{\partial}

\def\ket#1{ | #1 \rangle}
\def\bra#1{ \langle #1 |}
%
%
\def\CA{{\cal A}}		\def\CC{{\cal C}}
		\def\CF{{\cal F}}
		\def\CI{{\cal J}}

\def\CV{{\cal V}}		

\def\({ \left( }\def\[{ \left[ }
\def\){ \right) }\def\]{ \right] }
%


\def\IR{\relax{\rm I\kern-.18em R}}
\font\cmss=cmss10 \font\cmsss=cmss10 at 7pt
\def\IZ{\relax\ifmmode\mathchoice
{\hbox{\cmss Z\kern-.4em Z}}{\hbox{\cmss Z\kern-.4em Z}}
{\lower.9pt\hbox{\cmsss Z\kern-.4em Z}}
{\lower1.2pt\hbox{\cmsss Z\kern-.4em Z}}\else{\cmss Z\kern-.4em Z}\fi}
\def\inbar{\,\vrule height1.5ex width.4pt depth0pt}
\def\IB{\relax{\rm I\kern-.18em B}}
\def\IC{\relax\hbox{$\inbar\kern-.3em{\rm C}$}}
\def\ID{\relax{\rm I\kern-.18em D}}
\def\IE{\relax{\rm I\kern-.18em E}}
\def\IF{\relax{\rm I\kern-.18em F}}
\def\IG{\relax\hbox{$\inbar\kern-.3em{\rm G}$}}
\def\IH{\relax{\rm I\kern-.18em H}}
\def\II{\relax{\rm I\kern-.18em I}}
\def\IK{\relax{\rm I\kern-.18em K}}
\def\IL{\relax{\rm I\kern-.18em L}}
\def\IM{\relax{\rm I\kern-.18em M}}
\def\IN{\relax{\rm I\kern-.18em N}}
\def\IO{\relax\hbox{$\inbar\kern-.3em{\rm O}$}}
\def\IP{\relax{\rm I\kern-.18em P}}
\def\IQ{\relax\hbox{$\inbar\kern-.3em{\rm Q}$}}
\def\IGa{\relax\hbox{${\rm I}\kern-.18em\Gamma$}}
\def\IPi{\relax\hbox{${\rm I}\kern-.18em\Pi$}}
\def\ITh{\relax\hbox{$\inbar\kern-.3em\Theta$}}
\def\IOm{\relax\hbox{$\inbar\kern-3.00pt\Omega$}}


\def\mun{{\bf I}}

\def\oh{{1\over 2}}

\def\Ga{\alpha}
\def\Gd{\delta}\def\Ge{\epsilon}

\def\Gl{\lambda}


\def\mod{{\rm mod\,}}

\def\pd{\partial } 
\def\bra{\langle}\def\ket{\rangle}
\def\nind{\noindent}
\def\td{t_{\textstyle{.}}}

 \def\Che{Chebishev\ } 
 \def\LG{Landau-Ginsburg\ }
\def\ie{{\it i.e.\ }}

\Title{SPhT 93/147;  hep-th/9312209}
{{\vbox {
\centerline{On Dubrovin Topological Field Theories}
}}}

\bigskip

\centerline{J.-B. Zuber}\bigskip

\centerline{ \it Service de Physique Th\'eorique de Saclay
\footnote*{Laboratoire de la Direction des Sciences
de la Mati\`ere du Commissariat \`a l'Energie Atomique.},}
\centerline{ \it F-91191 Gif sur Yvette Cedex, France}

\vskip .2in

\noindent I show that the new topological
field theories recently associated by Dubrovin with each Coxeter group
may be all obtained in a simple way by a ``restriction'' of the
standard $ADE$ solutions. I then study the Chebichev specializations
of these topological algebras, examine how the Coxeter graphs and
matrices reappear in the dual algebra and mention the intriguing
connection with the operator product algebra of conformal field theories.
A direct understanding of the occurrence of Coxeter groups in that
context is highly desirable.

\Date{12/93\qquad\qquad Submitted for publication to
{\it Modern Physics Letters}}
%


\lref\Arn{V.I. Arnold, S.M. Gusein-Zade and A.N. Varchenko,
{\it Singularities of differentiable maps}, Birk\"auser, Basel 1985.}
\lref\VC{A.N. Varchenko and S.V. Chmutov, Funct. Anal. \& Appl.
{\bf 18} no 3 (1984) 171
[{\sl Funk. Anal. Priloz. {\bf 18 } (1983) 1}].}

\lref\DVV{R. Dijkgraaf, E. Verlinde and H. Verlinde, Nucl. Phys.
{\bf B352} (1991) 59; in {\it String Theory and Quantum Gravity},
proceedings of the 11990 Trieste Spring School, M. Green et al. {\it eds.},
World Sc. 1991.}
\lref\EW{E. Witten, Nucl. Phys. {\bf B 340} (1990) 281.
}
\lref\EY{T. Eguchi and S.-K. Yang, Mod. Phys. Lett. {\bf A5} (1990) 1693.}
\lref\CV{C. Vafa, Mod. Phys. Lett. {\bf A6} (1991) 337.}
\lref\DFLZ{P. Di Francesco, F. Lesage and J.-B. Zuber,
Nucl. Phys. {\bf B408} (1993) 600.}
\lref\DFZf{P. Di Francesco and J.-B. Zuber, J. Phys. {\bf A 26} (1993) 1441.}
\lref\VWLM{E. Martinec, {\it Phys. Lett.} {\bf B217} (1989) 431;
{\it Criticality, catastrophes and compactifications}, in
{\it Physics and mathematics of strings}, V.G. Knizhnik memorial volume,
L. Brink, D. Friedan and A.M. Polyakov eds., World Scientific 1990
 \semi C. Vafa and  N.P. Warner, {\it Phys. Lett.} {\bf B218} (1989) 51.}
\lref\LVW{W. Lerche, C. Vafa, N.P. Warner, {\it Nucl. Phys.} {\bf B324}
(1989) 427.}
\lref\Wa{N. Warner, {\it $N=2$ Supersymmetric Integrable Models and
Topological Field Theories}, to appear in the proceedings of the 1992
Trieste Summer School, hep-th/9301088. }
\lref\LW{W. Lerche and N.P. Warner, 
in {\it Strings \& Symmetries, 1991}, N. Berkovits, H. Itoyama et al. eds,
World Scientific 1992.}

\lref\Dubr{B. Dubrovin, Nucl. Phys. {\bf B 379} (1992) 627.}
\lref\Dub{B. Dubrovin,
{\it Differential Geometry of the space of orbits of a Coxeter group},
preprint hep-th/9303152.}
\lref\Kr{I. Krichever, Comm. Math. Phys. {\bf 143} (1992) 415. }

\lref\Cox{H.S.M. Coxeter and W.O.J. Moser, {\it Generators and Relations for
Discrete Groups}, Springer 1957 \semi
N. Bourbaki, {\it Groupes et Alg\`ebres de Lie}, chap. 4--6, Masson 1981.}
\lref\Hum{J.E. Humphreys, {\it Reflection Groups and Coxeter Groups}, Cambridge
Univ. Pr. 1990.  }
\lref\Meh{M.L. Mehta, {\it Basic sets of invariant polynomials for
finite reflection groups}, Comm. Alg. {\bf 16} (1988) 1083
\  and further references therein.}
\lref\JBZ{J.-B. Zuber, Phys. Lett. {\bf 176} (1986) 127. }

\lref\VP{V. Pasquier, J. Phys. {\bf A 20} (1987) 5707.}
\lref\PZ{V. Petkova and J.-B. Zuber, to appear.}
\lref\CF{P. Christe, PhD thesis, Bonn preprint IR-86-32\semi
P. Christe and R. Flume, Phys. Lett. {\bf B188} (1987) 219.}

\newsec{Restriction}
\noindent
In a topological field theory (TFT) the genus zero
3-point correlation functions
$C_{ijk}=\bra \phi_i \phi_j \phi_k\ket$ are sufficient to reconstruct all
the others. They are functions of deformation parameters $t_l$ and in
this article, we shall consider TFT's with a finite number of
fields and parameters, $i,j,k,l\in \{0,1,\cdots,n\}$. Then the $C$'s
must satisfy a set
of conditions \refs{\EW{--}\DVV}\ (the ``Witten-Dijkgraaf-Verlinde-Verlinde''
equations):
\eqna\Iao
$$\eqalignno{\bullet\quad&
\exists F(\td)\ {\rm such\ that}\
C_{ijk}= {\pd^3{F}\over \pd t_i\pd t_j\pd t_k} & \Iao a\cr
\bullet\quad&
\eta_{ij}= C_{0ij} \ {\rm is\ independent\ of\ the }\ t
{\rm 's\ and\ invertible:}\ \eta_{ij}\eta^{jk}=\Gd_i^k   &\Iao b\cr
\bullet\quad&
C_{ij}^{\ \ k}= \eta^{kl}C_{ijl} \ {\rm are\ the\ structure\ constants\
of\ an\ associative\ algebra
\ }\CA\ i.e.\cr
&  \qquad
C_{ij}^{\ \ k}C_{kl}^{\ \ m}= C_{il}^{\ \ k}C_{kj}^{\ \ m} & \Iao c\cr
\bullet\quad&
F(\td) \hbox{ is a quasi-homogeneous function of the }t{\rm 's . }
& \Iao d\cr }$$
The function $F(t_0,t_1,\cdots,t_n)$ is the {\it free energy} of the theory.
It encodes all the information about the $C$'s.

Dubrovin \Dubr\ has reintrepreted these conditions in a geometric,
coordinate invariant way, as defining a {\it Frobenius manifold}.
More recently  \Dub, he has shown how to associate a Frobenius manifold, hence
a solution to \Iao{}, with each finite Coxeter group.
Coxeter groups are linear groups generated by reflections in a real
Euclidean space $V$.
The finite Coxeter groups are classified \refs{\Cox{--}\Hum}~:
in addition to the Weyl groups of the simple
Lie algebras, $A_p$, $B_p$, $C_p$, (the two latter Coxeter groups
being identical), $D_p$, $E_6$, $E_7$, $E_8$, $F_4$ and
$G_2$, there are the groups $H_3$  and $H_4$ of reflections of the
regular icosaedron and of a regular 4-dimensional polytope, and
the infinite series $I_2(k)$ of the symmetry groups
of the regular $k$-gones in
the plane \refs{\Cox{--}\Hum}. In Dubrovin's work, the homogeneity degrees
of the variables $t_i$ and of $F$ are respectively $1-(d_i-2)/h$ and
$2+2/h$ where $h$ is the Coxeter number of $G$
and $d_i$ are the degrees of the $G$ invariant polynomials in the coordinates
of $V$ ($d_0=2$,\dots $d_n=h$).
Moreover, Dubrovin has proved a unicity theorem asserting that his solutions
are the only ones with these assignments of degrees, and he
conjectures that they are the only solutions with $F$ polynomial and
homogeneity degrees satisfying
\eqn\Iaoo{ 0< {\rm degree}\,(F)-2\le\,{\rm degree}\,(\td)\le 1\ .}

The solutions of type $ADE$ that Dubrovin finds reproduce what is
already well known as the minimal TFT's, obtained by twisting
and perturbing the minimal $N=2$ superconformal theories
(for explicit expressions of their free energies, see \DVV\DFLZ\ and
further references therein).

The purpose of this note is to show that the other solutions
may be obtained from the latter by a ``restriction'', 
thus obtaining in a
simple and explicit way their free energy $F$, and to point to
some 
properties of these additional solutions and connections with other problems.
This restriction amounts to setting some of the
$t$ parameters of an $ADE$ solution to zero, $t_i=0$ if $i\notin I$, and
to concentrating on the $C_{ijk}$, $i,j,k\in I\subset \{0,\cdots,n\}$.
To prove that such a restriction
still yields a solution to \Iao{}, the only thing to check is that
the restricted $C$'s form a subalgebra, namely
%
\eqn\Iap{C_{ij}^{\ \ k}(\td)|_{t_l=0, l\notin I} =0 \qquad \forall i,j\in I
,\quad \forall k\notin I .}
In other words, searching for restrictions amounts to looking for
subalgebras of some specialization (for some $t_i=0, i\notin I$)
of the $ADE$ topological algebras.

These restrictions come in two classes. The first is associated with
symmetries of some solution of $ADE$ type. Suppose that the free energy
of an $ADE$ solution is left
invariant by some group of transformations of its arguments
$t_l\to \omega_l t_l$. It
turns out that the symmetry groups are either $\IZ_2$ or $\IZ_3$. The
restriction consists in a projection on the invariant sector,
$I=\{j|\omega_j=1\}$.
Clearly condition \Iap\ is fulfilled since
$C_{ij}^{\ \ k}(\td)|_{t_l=0, l\notin I} $ is not invariant under the
action of the symmetry group. This procedure yields
the $B_n\equiv C_n$, $F_4$ and $G_2$ solutions. The other class
of restriction does not seem to be associated with any symmetry; this
is the way the $H_3$, $H_4$ and $I_2(k)$ solutions will be
obtained.

In the following, instead of labelling the $t$ parameters of the
$ADE$ solutions in a consecutive way, from $0$ to $n$, we index them
by the degree $-2$ of the associated invariant polynomial~: $d_i=i+2$.
The two labellings coincide in the $A$ cases. Through the restriction, we
shall see that the $t$'s of the non $ADE$ solutions inherit a labelling
consistent with this convention.

$\bullet$ Consider first the free energy of the $A_{2n+1}$ case.
It is known to be invariant under the following $\IZ_2$ action on the
$t$ parameters~: $t_k \to (-1)^k t_k$. It is thus consistent
to let all $t$'s of odd index vanish.
Then
\eqn\Ib{F_{B_{n+1}}
( t_0,t_2,\cdots, t_{2n})=F_{A_{2n+1}}( t_0,0, t_2,
\cdots,0, t_{2n})\ . }
%
In particular
\eqnn\Ibb
$$\eqalignno{F_{B_2}(t_0,t_2)&={t_0^2 t_2\over 2}+{t_2^5\over 60}& \Ibb\cr
F_{B_3}(t_0,t_2)&={t_0^2 t_4\over 2}+{t_0 t_2^2\over 2}+{t_2^3 t_4\over 6}+
{t_2^2 t_4^3\over 6}+ {t_4^7\over 210}\ .\cr}$$
Notice that this procedure of restricting oneself
to the even $t$'s  resembles that yielding the $D_{n+2}$ case from the
$A_{2n+1}$ one \DVV,\DFLZ. In the latter case, however, this is accompanied
by the introduction of a new parameter $t'_n$ and one is really
dealing with an {\it orbifold} of $A$ rather than a restriction.

$\bullet$ The $E_6$ free energy
$F_{E_6}(t_0,t_3,t_4,t_6,t_7,t_{10})$ is invariant under
the same $\IZ_2$ action $t_k \to (-1)^k t_k$.
The $F_4$ solution may thus be obtained
from the $E_6$ one~:
\eqnn\Ic
$$\eqalignno{F_{F_4}(t_0,& t_4,t_6,t_{10}) = F_{E_6}(t_0,0,t_4,t_6,0,t_{10})
& \Ic \cr
&= t_0t_4t_6+{t_0^2t_{10}\over 2}+{t_4^3t_{10}\over 6}+{t_4t_6^2t_{10}^3
\over 6}+{t_6^4t_{10}\over 12}+{t_4^2t_{10}^5\over 60}+
{t_6^2t_{10}^7\over 252}+{t_{10}^{13}\over 185328}\ .\cr
}$$

$\bullet$ In a similar way, the $D_4$ free energy written as
\eqn\Icc{F_{D_4}={t^{2}_{0}\,t_{4 }\over{2}}
+{t_{0}(t^{2}_{2}-t^{'2}_{2})\over{2}}
 +{t^{3}_{4}\over{6}} ({t^{2}_{2} -t^{'2}_{2}})
 +t_{4} \,({t^{3}_{2} \over{6}} +{{t_{2}\,t^{'2}_{2}}\over{ 2}} )
+{{t^{7}_{4}}\over{210}} }
is invariant under the $\IZ_3$ symmetry $t_0\to t_0$, $t_4\to t_4$,
$t_2\pm t'_2\to \exp \pm 2i\pi/3 \ (t_2\pm t'_2)$. Projecting again
onto the invariant sector, the $G_2$ solution is obtained
from the $D_4$ one,
\eqn\Id{F_{G_2}(t_0,t_4)=F_{D_4}(t_0,t_2=0,t'_2=0,t_4)\ .}
Explicitly
\eqn\Ie{F_{G_2}(t_0,t_4)={t_0^2 t_4\over 2}+{t_4^7\over 210}\ .}

\bigskip

We now come to the second class of restrictions.

$\bullet$ Let us consider the $E_8$ solution
 $F(t_0,t_6,t_{10},t_{12},t_{16},t_{18},t_{22},t_{28})$
and let $t_6=t_{12}=t_{16}=t_{22}=0$. It is a tedious but straightforward
exercise to check on the explicit expression given in \DFLZ\
that the terms in ${\pd^2 F\over \pd t_i \pd t_j}$, $i,j\in I=\{0,10,18,28\}$
that contain a $t_k\notin I$ are at least quadratic in the $t_l$, $l\notin I$,
thus proving \Iap. 
Only 12 of the original 140 terms of $F_{E_8}$ survive ! and
we obtain in this way the $H_4$ solution of Dubrovin
\eqnn\Iee
$$\eqalignno{F_{H_4}(&t_0,t_{10},t_{18},t_{28})=
t_0\,t_{10}\,t_{18}
+{{t_0^{2}\,t_{28}}\over{2}}
+t_{18}^{2}\, {{t_{28}^{19}}\over{1539000}} \cr
& +{{t_{18}^{3}\,\over{6}}\left({{t_{28}^{13}}\over{1800}}
+{{t_{18}\, t_{28}^{7}}\over{20}}
+{{3\,t_{18}^{2}\,t_{28}}\over{10}} \right)}
+{{t_{28}^{31}}\over{ 245764125000}}
& \Iee\cr &
+t_{10}^{2}\, \left({{t_{28}^{11}}\over{4950}}
+{{t_{18}\,t_{28}^{5} }\over{20}}
+{{t_{10}\, t_{28}}\over{6}} \right)
+t_{10}\,t_{18}^{2}\,\left({{t_{28}^{9} }\over{360}}
+{{t_{18}\,t_{28}^{3}}\over{6}} \right)
\ . \cr}$$

\def\bt{{\bf t}}
$\bullet$ In a similar way, the $H_3$ solution may be obtained
from the $D_6$ one.
It was shown in \DFLZ\ that the $D_6$ solution
is also related by slight changes of
parametrization to a $SU(3)$ related solution. The restriction that yields
the $H_3$ case is simpler to express in terms of the latter.
One finds, denoting with boldface $\bt$'s and $\tau$
 the $D_6$ parameters in the notations of \DFLZ\
(and with $a^4=-4$)\foot{
There are unfortunate misprints in the last four lines of
Appendix C of \DFLZ: one should read $t_{11}=-a^3 \bt_3$; $t_{10}=2\bt_1$;
$t_{00}=2a\bt_0$.}
\eqnn\If
$$\eqalignno{F_{H_3}(t_0,t_4,t_8)
&= F_{SU(3)_2}(t_{00}=t_0,t_{10}=0,t_{01}=0,t_{20}=t_4,t_{11}=0,t_{02}=t_8)
 \cr
&= a^7 F_{D_6}(\bt_0={1\over 2a}t_0,\bt_1=0,\bt_2=-{1\over a^3}t_4,
\bt_3=0,\bt_4=-{1\over a}t_8,\tau=-{1\over 2a}t_4) \cr
&= {t_0^2 t_8\over 2}+{t_0 t_4^2\over 2}
+{t_4^3 t_8^2\over 6}+{t_4^2 t_8^5\over 20}+{t_8^{11}\over 3960}\ .&\If
 \cr }$$

$\bullet$ Finally, I claim that one may set all $t$'s but $t_0$ and $t_n$
equal to zero in the $A_{n+1}$ free energy and get a consistent solution.
This will be the $I_2(n+2)$ solution of \Dub. To prove this, we recall
that {\bf (i)},
in $F_{A_{n+1}}$, the  variable $t_i$ appears in terms that are
at most $(i+3)$-linear in the $t$'s \DVV, and that {\bf (ii)}, $F$ is a
quasihomogeneous function of degree $2+2/(n+2)$ of the $t$'s if $t_i$ is
assigned the degree $1-i/n+2$. The only terms in $F$ involving $t_0$ are
$\oh t_0\sum_{k=0}^n ( t_k t_{n-k})$: they generate terms that satisfy \Iap.
Let us now look at the others. They would violate \Iap\ if and only if
they are
of the form $t_n^p  t_k$, $p\ge 2$, $1\le k\le n-1$. This, however,
has degree $1+(2p-k)/n+2=2+2/n+2$, whence $2p=n+4+k$
and by the above observation {\bf (i)}, $p< k+1$, which leads to the
contradiction $k\ge n$. This completes the proof of the
consistency of this restriction, leading to the $I_2(n+2)$ solution
\eqnn\Ig
$$\eqalignno{F_{I_2(n+2)}(t_0,t_n )&=F_{A_{n+1}}(t_0,0,\cdots,t_n) \cr
&= {t_0^2 t_n\over 2} +{t_n^{n+3}\over (n+1)(n+2)(n+3)} &\Ig
\ .}$$

\bigskip

In all cases, the labelling of the $t$'s is such that these labels
take their values in the
``exponents'' minus one  of the corresponding Coxeter group $G$, \ie\
the degrees minus two of a basis of the ring of invariant polynomials
\refs{\Cox{--}\Hum}~: $d_i=i+2$.
The homogeneity degrees of the variables $t_i$ and of $F$
are respectively $1-i/h$ and $2/h+2$ where $h$ is the Coxeter number of $G$.
In view of the unicity theorem of \Dub, this justifies that we have
found the desired solutions.

Also we check that the expressions \Ibb, \Id\ and \Ig\
are consistent with the well known identifications of Coxeter groups
$A_2\equiv I_2(3)$, $B_2\equiv I_2(4)$ and $G_2\equiv I_2(6)$.
In fact, all these expressions are defined up to a change of
normalization of $F$ and of each $t$.

Note that all the restrictions of $ADE$ cases that we have
found respect the $t$ of lowest label 0 --- as expected--- and
highest $h-2$, hence preserve
the Coxeter number $h$. This is because they have to
respect the pairing between 0 and $h-2$ in the metric
$$ \eta_{0i}=C_{00i}= \Gd_{i,h-2}\ .$$

At this stage, it is not clear that we have exhausted all possibilities of
restriction (with $0,h-2\in I $),
although Dubrovin's conjecture asserts so.
We are going to prove it by specializing our algebras in a
definite way, namely by letting all $t$ vanish but $t_{h-2}$, the one
with the smallest homogeneity (the least relevant coupling
in physical terms). Recall that $t_0$ (which is of homogeneity
degree 1) can appear at most
quadratically in $F$ (of degree $<3$),
hence doesn't appear in the $C$'s. In this specialization, the $C$'s
depend only on $t_{h-2}$ in an homogeneous way, hence $t_{h-2}$
may be taken equal to 1 with no loss of generality.
In the simplest case of the $A_{k+1}$ topological algebra, this
specialization is known to reproduce the fusion algebra of $\widehat{su}(2)_k$
that has a polynomial representation in terms of \Che polynomial.
We thus call this specialization in general 
the \Che specialization.


\newsec{The \Che specialization}
\nind
This specialization to all $t_i=0$ but $t_{h-2}=1$ of the $C$ algebra
is known to enjoy many nice properties, in the $ADE$ cases \LW\DFLZ.
After studying all the subalgebras of the $ADE$ algebras, we shall extend these
properties to them and present a curious connection with operator
product expansions (OPE) of conformal field theories (CFT).

\subsec{The subalgebras of the $ADE$ algebras}
\nind We shall now prove that the only subalgebras containing the
generators of smallest and largest 
label $0$ and $h-2$ of the $ADE$ algebras are those associated
by the construction of the first section to the other Coxeter groups.

First consider the $A_{k+1}$ case; as recalled above, it is isomorphic
to the $\widehat{su}(2)_k$ fusion algebra, \ie
%
\eqn\IIa{\phi_i \phi_j=\sum_{l=|i-j|,|i-j|+2,\cdots, {\rm inf}
(i+j,2k-i-j)} \phi_l\ .}
Suppose that we have a subalgebra containing $\phi_0$, and let $i$ be the
smallest non zero label such that $\phi_i$ belongs to the subalgebra.
Consider
\eqn\IIb{\phi_i \phi_i=\phi_0+\cdots +\phi_{{\rm inf}(2i,2k-2i)}\ .}
Either ${\rm inf}(2i,2k-2i)=0$, \ie $i=k$, and this is what we
call the $I_2(k+2)$ algebra. Or $\phi_2$ belongs to the sum \IIb, hence
$i\le 2$. If $i=1$, we are back to the original $A_{k+1}$ algebra, whereas
$i=2$ corresponds to the subalgebra
$\{\phi_0,\phi_2,\cdots,\phi_{2[{k\over 2}]}\}$; this subalgebra
satisfies the extra requirement to contain  the generator $\phi_k$
of largest Coxeter exponent only if $k$ is even~: we then get the
$B_{k+1}$ solution.

The case of the $D_{2p+2}$ algebra is discussed along similar lines. This
algebra is known to have $2p+2$ generators 
$\{\phi_0,\phi_2,\cdots,\phi_{4p},\alpha\}$ satisfying the relations
\IIa\ together with
\eqnn\IIc
$$\eqalignno{\alpha \phi_{2l}&= (-1)^l\alpha  & \IIc \cr
\Ga \Ga &= \phi_{0}-\phi_{2}+\cdots +\phi_{4p}\cr }$$
[This should not be confused with the $D_{2p+2}$ {\it extended} algebra
generated by the combinations $\Phi_l=\oh(\phi_{2l}+\phi_{4p-2l})$,
$l=0,1\cdots p-1$ and $\Phi_p^{\pm}=\oh(\phi_{2p}+\Ge \Ga)$,
$\Ge^2=(-1)^{p}$~: see for example \DFZf].
Beside the obvious subalgebras $B_{2p+1}=\{\phi_0 ,\phi_2,\cdots \phi_{4p}\}$
and $I_2(4p+2)=\{\phi_0,\phi_{4p}\}$ already encountered, the unique
other possibility is to generate the subalgebra by a combination of
two generators of $D$ of same degree, namely $\oh(\phi_{2p}+ \Ge \Ga)$.
This turns out to be consistent only for $p=2$, $\Ge^2=1$,
and we have the algebra $H_3=\{\phi_0,\oh(\phi_4\pm  \Ga), \phi_8\}$.

This possibility doesn't exist for the $D_{2p+1}$ cases, for which we
conclude that there is no other subalgebra.

Finally the three exceptional cases are handled case by case, with the
anticipated result that the only subalgebras (other than $I_2$ and containing
$\phi_{h-2}$) are  those corresponding to $F_4$ and $H_4$, in $E_6$ and
$E_8$ respectively.


\subsec{The dual algebras}
\nind In \LW, it was noticed that the $C$ matrices
of the $ADE$ \Che\ specializations have, after an appropriate change of
basis, the same eigenvectors as the $ADE$ Cartan matrices. That
was studied more systematically in \DFLZ, where it was shown that
the matrices $(C_i)_j^{\ k}=C_{ij}^{\ \ k}$ can be made normal
(\ie commuting with their transpose) by a diagonal change of basis, and
that their normal form $M_i$  may be written in terms of the
orthonormalized eigenvectors $\psi^{(i)}_a$ as follows
\eqnn\IIf
$$\eqalignno{C_{ij}^{\ \ k}&= {\rho_i \rho_j\over \rho_k}
M_{ij}^{\ \ k}\cr
M_{ij}^{\ \ k}&= \sum_a {\psi^{(i)}_a\psi^{(j)}_a\psi^{(k)\, *}_a
\over \psi^{(0)}_a}\ .&\IIf  \cr}$$
Here $\psi^{(i)}_a$ denotes the $a$-th component
of the $i$-th eigenvector of the Cartan matrix, or equivalently,
since we are still considering the simply laced $ADE$ cases,
of the adjacency matrix $G_a^{\ b}$ of the Dynkin diagram. Thus the
label $a$ runs over the vertices of the Dynkin diagram, and the
label $i$ over the Coxeter exponents (minus 1), in accordance
with our previous conventions.

In \DFLZ, it was further observed that it is interesting to also
look at the {\it dual} algebra structure generated by the  matrices
$N_a$ of matrix elements
\eqn\IIg{N_{ab}^{\ \ c}= \sum_i {\psi^{(i)}_a\psi^{(i)}_b\psi^{(i)\, *}_c
\over \psi^{(i)}_{0}} \ .}
This is again an associative algebra, with the matrix $N_{0}=\mun$, the
identity matrix. Eq. \IIg\ assumes that
there exists at least one vertex labelled $0$ such that
all the $\psi^{(i)}_{0}$ are non vanishing. If we may take such a
point among the extremal vertices of the diagram, \ie those
connected with only one other vertex denoted $f$, then it is
easy to see that $\psi^{(i)}_{f}/\psi^{(i)}_{0}=\Gl^{(i)}$
is the $i$-th eigenvalue of the adjacency matrix $G$, hence
$N_{f}=G$. In fact, one may always choose the point $0$ as
the end point of the long branch of the Dynkin diagram, except for the
$D_{{\rm odd}}$ cases, where one has to take it as the end point of
one of the short branches.

To summarize, the dual algebra is generated by the identity matrix
$N_{0}$ and the adjacency matrix of the Dynkin diagram $N_{f}=G$,
all the matrices $N_a$ thus have integral entries and,
depending on the case (and the choice of the vertex $0$), these integers
are or are not all non negative (cf.\DFLZ).

\medskip

It is now natural to wonder if these properties extend to the subalgebras
labelled by other Coxeter groups that we have discussed (still in the
\Che specialization). By a case by case analysis, one may convince oneself
that
\item{$\star$}\ for all the $B_n,F_4,G_2,H_3,H_4,I_2$ cases, the $C$ matrices
may again be brought to a normal form by a diagonal change of basis;
\item{$\star$}\ the normal form $M_i$ of $C_i$ has only non negative
entries;
\item{$\star$}\ in contrast with the $ADE$ cases, the matrices of the
dual algebra are no longer all with integral entries; the dual algebra,
however,
is generated by the identity matrix and the matrix $N_f$
\eqnn\IIh
$$\eqalignno{(N_f)_{a}^{\ a}&=0 & \IIh\cr
(N_f)_{a}^{\ b}&=2 \cos{\pi\over m(a,b)}\qquad a\ne b \ ,\cr}$$
where the integer $m(a,b)=m(b,a)$ takes a value different from
2 or 3 (hence $N_a^{\ b}\ne 0,1$) only for one pair of vertices $(a,b)$;
$m=4$ for the $B$ and $F_4$, $m=5$ for the $H_3,\ H_4$ cases
(for $I_2$ see below).

\noindent
In fact one recognizes in these $N$ matrices essentially
the ``Coxeter matrices'' of the corresponding ``Coxeter graphs''. Let us
recall the definitions, following  \Hum. 
We consider an unoriented graph, the edges $(a,b)$ of which are labelled
by integers $m(a,b)=m(b,a) \ge 3$. It is convenient to extend this
function $m$ to all pairs of vertices of the graph~:
$m(a,a)=1$, $m(a,b)=m(b,a)=2$ if $a$ and $b$ are not neighbours on the graph.
These integers are tabulated for the relevant graphs in Table I. To such
a  graph, we associate the Coxeter matrix
\eqn\IIha{ \Gamma_{ab}=-\cos{\pi \over m(a,b)}\ .}
%
For the Weyl-Coxeter groups, this is
(up to a factor 2) the ``normalized'' form of the Cartan matrix
\eqnn\IIi
$$\eqalignno{ \CC_{ab}&=2{\bra \Ga_a,\Ga_b\ket\over \bra\Ga_b,\Ga_b\ket}\cr
&= {|\Ga_a|\over|\Ga_b|}\, 2\cos(\Ga_a,\Ga_b)& \IIi\cr
&=-{|\Ga_a|\over|\Ga_b|}\, 2\cos\({\pi\over m(a,b) }\)\ .\cr }$$
Now the matrix $N_f$ of \IIh\ reads
\eqn\IIj{N_f=2(\mun - \Gamma) \ .}
%
\bigskip
\vbox{
\ifx\answ\bigans\hrule height 1pt depth 0pt width 165truemm
\else\hrule height 1pt depth 0pt width 12truecm\fi
$$\eqalignno{
B_n &\quad
\bullet\!\!{{}\over{\qquad}}\!\!\bullet\!\!{{}\over{\qquad}}\!\!\bullet
\cdots\cdots
\bullet\!\!{{4}\over{\qquad}}\!\!\bullet \cr
F_4 & \quad
\bullet\!\!{{}\over{\qquad}}\!\!\bullet\!\!{{4}\over{\qquad}}\!\!\bullet\!\!
{{}\over{\qquad}}\!\!\bullet \cr
H_3 & \quad
\bullet\!\!{{}\over{\qquad}}\!\!\bullet\!\! {{5}\over{\qquad}}\!\!\bullet \cr
H_4 &\quad
\bullet\!\!{{}\over{\qquad}}\!\!\bullet\!\!{{}\over{\qquad}}
\!\bullet\!\! {{5}\over{\qquad}}\!\!\bullet \cr
I_2(p) & \quad
\bullet\!\! {{p}\over{\qquad}}\!\!\bullet \cr
}$$
%
\smallskip
\centerline{Table I : Coxeter graphs}
\centerline{ The well known $ADE$ diagrams are not represented here;
neither is $G_2=I_2(6)$;}
\centerline{
$m(a,b)=m(b,a)=3$ unless otherwise specified above the
edge $(a,b)$.   }
\medskip
\ifx\answ\bigans\hrule height 1pt depth 0pt width 165truemm
\else \hrule height 1pt depth 0pt width 12truecm\fi
} 
%
%
%
%
\bigskip

The only exception to our discussion is the $I_2(n+2)$ case, which is
somehow degenerate~: the dual algebra is generated by the two-by-two
matrices $\mun=\pmatrix{1&0\cr0&1\cr}$ and $\pmatrix{0&1\cr1&0\cr}$;
the later matrix is a multiple of the Coxeter
matrix $\pmatrix{0&\cos{\pi\over n+2}
\cr\cos{\pi\over n+2}&0\cr }$ 
attached to the $I_2(n+2)$ graph, and our dual algebra does not ``see''
the factor $\cos{\pi \over n+2}$.

\medskip

To summarize, except maybe in the $I_2$ case, the dual algebra exhibits the
most natural generalization of the property observed in the simply laced
case, namely it is generated by the Coxeter matrices of the relevant
diagrams.
It is quite curious to see the emergence of these Coxeter matrices
in this problem. Recall that the $A\cdots
I$ Coxeter graphs solve the following
problem~: they are the unique graphs such that the associated
Coxeter matrices \IIha\ are positive definite, or equivalently, such
that the matrix $N_f$ of \IIj\ has all its eigenvalues
smaller than 2. It would be quite
interesting to see directly why this property is requested of the
dual algebra.

%
\subsec{A connection with OPE of CFT}
\nind
We finally turn to a remarkable fact that gives another perspective
to (the \Che specializations of) these topological algebras~:
their connection with operator product expansions in WZW or minimal
conformal field theories. That the $C$ algebra of the $A_{k+1}$ case
yields the fusion algebra of the $\widehat{su}(2)$ theories has been
recalled above. This is already quite astonishing, since it relates an
 algebra associated with a ``massive'' topological field theory
(or its perturbed $N=2$ partner) with a similar
structure in a {\it conformal}
theory. Moreover, this result has a non trivial extension to the other
simply laced cases $D$ and $E$. This fact, already
anticipated a long
time ago by Pasquier \VP, has been put in a quantitative form
in a collaboration with V. Petkova and will be presented in detail
elsewhere \PZ. In short, the normalized form $M_i$ of the $C_i$ matrices
yields the ratios of the structure constants of spinless operators in
the non diagonal $D$ or $E$ WZW $\widehat{su}(2)$ theories over the
same structure constants evaluated in the $A$ theory.

One would like to summarize the findings of this section by stating that
the subalgebras of the OPE of spinless fields of
$\widehat{su}(2)_k$ theories that contain
both the identity and the field of largest label  (twice the isospin)
 $i=k$ are classified
by Coxeter groups\foot{The issue of subalgebras of the OPE was also
addressed several years  ago by Christe and Flume \CF; they didn't, however,
impose the condition that $\phi_k$ should be kept in the subalgebra.}.
This is, unfortunately, not quite correct, as the spinless operators
of the WZW or minimal theories do  not form a closed algebra~:
the OPE of two such fields may involve non zero spin fields, and the
relevant structure constants do  not seem determined in a similar
algebraic way. The more correct (and cumbersome)
statement is therefore that the projection on
spinless fields of the subalgebras of the OPE of
$\widehat{su}(2)$ theories that contain
both the identity and the field of largest label $i=k$ are classified
by Coxeter groups.


\newsec{Discussion}

%
\noindent The $ADE$ solutions are known to admit a ``\LG picture'', which
means that the algebra $\CA$ with
structure constants $C_{ij}^{\ \ k}$ has a polynomial representation
\refs{\VWLM\LVW{--}\CV}
$$ \CA \equiv {\IC(x,y,\cdots)\over \CI(\pd_x W,\pd_y W,\cdots)}$$
where $W(x,y,\cdots;\td)$ is the ``\LG potential'', $\IC(x,y,\cdots)$
is the ring of complex polynomials in the indeterminates $x,y,\cdots$
and $\CI(\pd_x W,\cdots)$ is the ideal generated by the derivatives of $W$.
 In other words,
the representation is provided by the multiplication of the
polynomials $p_i(x,y,\cdots)=\pd W/\pd t_i$
$$ p_i p_j =C_{ij}^{\ \ k} p_k \qquad \mod \pd_x W,\ \pd_y W,\cdots \ .$$
This approach is deeply related to the theory of singularities~:
$W(x,y,\cdots;\td)$ is the versal deformation of the singularity
$W(x,y,\cdots;0)$ with the $t$'s the flat coordinates.

Our construction of the other, non $ADE$, Coxeter solutions to \Iao{}\
was based on the restriction procedure. This is a familiar
idea in singularity theory. There \Arn, it is known that
by quotient of an $A_{2n+1}$, $E_6$ or $D_4$ singularity by a
discrete symmetry,
one obtains a ``boundary singularity'' labelled by the $B$, $F_4$ or
$G_2$ Dynkin diagram. Moreover, the $H_3$ and $I_2$ cases have also
received a similar treatment \VC; it would be interesting
to see if the present construction of $H_3,\ H_4$ and $I_2$
by restriction suggests other possibilities.

\bigskip

In fact the same procedure has also been encountered in a different
context, that of modular invariant or sub-modular invariant partition
functions $Z$. The partition function of a  conformal
field theory on a torus encodes its field 
content that must be consistent with
(i) the operator product algebra, (ii) modular invariance. It may sometimes
be possible to find 
restrictions of this field content that are still consistent with a
closed operator product algebra but only
satisfy invariance under a subgroup of the modular group. For example,
it is known that modular invariant partition functions
of $N=2$ superconformal
(or affine $\widehat{su}(2)$ ) theories are classified according to an
$ADE$ scheme. One may verify that the fusion algebras of the $A_{2n+1}$
and $E_6$ cases admit a $\IZ_2$ symmetry.
Projecting on the even sector produces a partition function that is
invariant under a subgroup of the modular group \JBZ.
For example for $E_6$
$$Z_{E_6}=|\chi_0+\chi_{6}|^2+|\chi_3+\chi_{7}|^2+|\chi_4+\chi_{10}|^2\ ,$$
that is modular invariant, gives rise to
$$ Z_{F_4}=|\chi_0+\chi_{6}|^2+|\chi_4+\chi_{10}|^2$$
that is invariant under the subgroup $\hat\Gamma_0(2)$
of modular transformations $\pmatrix{a&b\cr c&d\cr}\equiv
\pmatrix{\pm 1 & * \cr0 &\pm 1\cr}\,\mod 2$.
Likewise, the modular invariant labelled $D_4$ leads under a $\IZ_3$ quotient
to a submodular invariant that we may label by $G_2$.
As we have seen in the first section,
the existence of a symmetry group is not required; the only condition
is that the operator product algebra of the restricted theory be closed.
We thus have additional cases corresponding to the $H_3$, $H_4$ and
$I_2$ groups:
$$\eqalignno{Z_{H_3}&=|\chi_{0}+\chi_8|^2+|\chi_{4}|^2   \cr
Z_{H_4}&=|\chi_{0}+\chi_{10}+\chi_{18}+\chi_{28}|^2 \cr
Z_{I_2(n+2)}&=|\chi_{0}|^2+|\chi_{n}|^2 \cr }$$
For each of these submodular invariants, the level of the affine
$\widehat{su}(2)$ algebra is read off the highest label, resp. 8,28 and $n$.
Just as in the end of last section, the role of the operators with
non zero spin is not clear, and one might for example
write as well $Z'_{H_3}=|\chi_{0}|^2+|\chi_8|^2+|\chi_{4}|^2 $.
This ambiguity notwithstanding, it appears that there is a
connection between these submodular invariants and Coxeter groups.

In this context of modular invariants,
we thus see that the $ADE$ and non $ADE$ situations
are not exactly on the same footing. The former describe sound
and consistent theories, whereas the latter describe some projection
thereof. As far as we can see,
this does not invalidate the TFT's associated with the latter,
but only reminds us that these cannot be obtained by a simple twisting
of a consistent, modular invariant, $N=2$ theory.

\bigskip

Finally, I return to the most intriguing and challenging point
already mentionned at the end of the previous section~:
the projections on spinless fields of consistent
operator product algebras of $\widehat{su}(2)$ theories containing the
field of largest isospin are classified by
Coxeter groups. If this could be established directly, independently of the
$ADE$ classification of modular invariants, it might offer an alternative
route to the latter. One would first classify the consistent operator
algebras as $AB\cdots I$ and then retain among them only the $ADE$
that give rise to a modular invariant partition function.

So we are left with the two questions
\item{$\star$}
Why are Coxeter groups involved in the operator algebra~?
\item{$\star$} What would be the generalization of this correspondence for
higher rank current algebras~?

\bigskip

{\bf Ackowledgements :} It is a pleasure to thank B. Dubrovin for
several very informative (electronic) conversations. I owe also
some useful observations to P. Di Francesco and V. Petkova.

\listrefs

\end